\documentclass[twocolumn,showpacs,showkeys, superscriptaddress,10pt]{revtex4-2}
\usepackage{amsmath,amssymb,graphicx, subfigure, hyperref, braket, float}
\usepackage{blindtext}

\counterwithout{figure}{section}
\usepackage[export]{adjustbox}
\graphicspath{{figure/}}
\usepackage[T1]{fontenc}
\usepackage{ae,aecompl}
\usepackage{graphicx, xcolor}
\begin{document}

\title{Intrinsic decoherence effects on correlated coherence and quantum discord in XXZ Heisenberg model}

\author{Zakaria Dahbi}
\email{zakaria\_dahbi2@um5.ac.ma}
\address{Lab of High Energy Physics-Modeling and Simulation, Faculty of Sciences, \\Mohammed V University in Rabat, 4 Avenue Ibn Battouta B.P. 1014 RP, Rabat, Morocco}

\author{Mansoura Oumennana}
\email{oum.mans@gmail.com}
\author{Mostafa Mansour}
\email{mostafa.mansour.fpb@gmail.com}
\address{Laboratory of High Energy Physics and Condensed Matter, Department of Physics, Faculty of Sciences Ain Chock, Hassan II University, P.O. Box 5366 Maarif, Casablanca 20100,  Morocco}

\begin{abstract}
Spin qubits are at the heart of technological advances in quantum processors and offer an excellent framework for quantum information processing. This work characterizes the time evolution of coherence and nonclassical correlations in a two-spin XXZ Heisenberg model, from which a two-qubit system is realized. We study the effects of intrinsic decoherence on coherence (\textit{correlated coherence}) and nonclassical correlations (\textit{quantum discord}), taking into consideration the combined impact of an external magnetic field, Dzyaloshinsky-Moriya (DM) and Kaplan-Shekhtman-Entin-Wohlman-Aharony (KSEA) interactions. To fully understand the effects of intrinsic decoherence, we suppose that the system can be prepared in one of the two well-known extended Werner-like (EWL) states. The findings show that intrinsic decoherence leads the coherence and quantum correlations to decay and that the behavior of the aforementioned quantum resources relies strongly on the initial EWL state parameters. We, likewise, found that the two-spin correlated coherence and quantum discord; become more robust against intrinsic decoherence depending on the type of the initial state. These outcomes shed light on how a quantum system should be engineered to achieve quantum advantages.
\end{abstract}

\keywords{intrinsic decoherence, spin qubits, quantum discord, correlated coherence}.

\maketitle

\section{Introduction}
Studying solid-state physical systems under multiple interactions has recently attracted particular focus. The unprecedented possibility of taking advantage of these systems has opened the door to constructing new quantum-based technologies \cite{dowling2003quantum}. Quantum superposition and entanglement \cite{einstein1935can} are two surprising aspects of the quantum theory and are found to be imperative resources for achieving speedup in information processing \cite{nielsen2002quantum} and for realizing several non-local tasks and classically unattainable applications \cite{ekert1992practical, kim2001quantum, mansour2020quantum, cruz2022quantum} within the realm of classical physics. Further, it has been demonstrated that even when entanglement is lost, quantum information processing tasks may still be performed in the case of particular mixed states due to the presence of nonclassical correlations beyond entanglement \cite{adesso2016measures, melo2017quantum}. On the other side, decoherence \cite{schlosshauer2019quantum} effects present a grave challenge to the beneficial applications of quantum mechanics because it prevents the retention and controllable handling of qubits.

Previously many studies focused on measuring and characterizing quantum resources in the framework of intrinsic and standard decoherence models in different quantum systems \cite{yin2022markovian, essakhi2022intrinsic, mohamed2020quasi, mohamed2022measurement, chaouki2022dynamics, hashem2022bell, mansour2020entanglement, haddadi2022measurement, maleki2017entanglement, wu2017intrinsic, li2021nonlocal, el2022dynamics, mohamed2022intrinsic, Mansour_2021}. Although many methods were introduced to measure quantum systems, assessing the quantumness of multipartite systems is still a challenging task \cite{adesso2016measures, schlosshauer2005decoherence, coopmans2022optimal}. For specific quantum systems, quantum discord (QD)  \cite{ollivier2001quantum} was the first quantifier introduced to capture nonclassical correlations. QD measures the difference between total correlations and classical correlations in a quantum system. However, it is a strenuous measure to calculate, and analytical expressions of QD were only obtained for two-qubit states \cite{luo2008quantum, Hender} and quantum X-states ( see for instance the works \cite{Li, Chen, haddadi, baba} and references therein). A rigorous mathematical characterization on the computational difficulties of quantum discord is provided in \cite{huang2014computing}, demonstrating that computing quantum discord is an NP-complete problem. Quantum correlations have been studied in many quantum systems, such as Heisenberg spin chain models  (for instance, spin-1 Heisenberg chains \cite{malvezzi2016quantum}, anisotropic spin-1/2 XY chain in transverse magnetic field \cite{mofidnakhaei2018quantum}) and quantum dot systems such as two coupled double quantum dots systems \cite{filgueiras2020thermal, elghaayda2022local} and double quantum dot system with single electron under Rashba interaction \cite{ferreira2022thermal} ). Recently, considerable attention has been dedicated to studying the influence of the Dzyaloshinsky-Moriya (DM) interaction and the Kaplan-Shekhtman-Entin-Wohlman-Aharony (KSEA) interaction on the quantum features of specific quantum systems \cite{oumennana2022quantum, khedif2021thermal, xie2022enhancing, oumennana2022}.

Quantum coherence is another quantum feature that one should deal with. This latter emanates from the superposition principle of quantum states and is a valuable resource to be preserved due to its pivotal role in quantum information processing \cite{pan2017complementarity} and quantum thermodynamics \cite{narasimhachar2015low}. It was indeed proven that long-lasting quantum coherence is vital for overcoming classical limitations of measurement accuracy in quantum metrology \cite{pires2018coherence}. Moreover, growing interest is accorded to the role played by quantum coherence in some biological mechanisms \cite{lambert2013quantum}, such as photosynthesis \cite{scholes2011coherence} and bird navigation \cite{ritz2011quantum}. Similarly to quantum correlations and entanglement, several quantifiers were introduced to capture quantum coherence in quantum systems appropriately, \textit{e.g.} relative entropy of coherence and $l_1$-norm of coherence \cite{baumgratz2014quantifying} as well as intrinsic randomness \cite{yuan2015intrinsic}.

In this article, we investigate the evolution of correlated coherence and quantum correlations quantified by quantum discord in a two-qubit XXZ Heisenberg spin chain under intrinsic decoherence effects with an applied magnetic field, KSEA and DM interactions using the Milburn's decoherence model \cite{milburn1991intrinsic}. We mainly focus on how the Hamiltonian parameters and the initial state affect the dynamical behavior of correlated coherence and quantum discord. We look at two different extended Werner-like states to show how the choice of the initial state is essential to suppress the effects of specific interactions.

The rest of this work is as follows. In Sec. \ref{sec2}, we give a short overview of the quantum resources indicators used in this study. The Hamiltonian of the considered physical model is introduced in  Sec. \ref{sec3}, and  the evolved density matrix corresponding to the system is derived for  the parameterized initial state considered. In Sec. \ref{sec4} we present the results obtained for the dynamics of correlated coherence and quantum discord under the influence of intrinsic decoherence  with the combined impact of KSEA and DM interactions. Finally, we conclude the main findings from the proposed model  in  Sec. \ref{sec5}.

\section{Quantum information indicators \label{sec2}}
This part defines the correlated coherence, used as a quantum coherence quantifier, and the quantum discord operated to quantify nonclassical correlations.

\subsection{Correlated coherence}
Quantum coherence is a basis-dependent property of a quantum state that arises from the superposition of system states reflected by the non-diagonal components of the density matrix on a given basis. Here, we consider a conveniently computable coherence measure, specifically the $l_1$--norm of coherence \cite{baumgratz2014quantifying}. The $l_1$-norm of coherence is a trustworthy key indicator that meets the requirements of a good coherence measure and can be obtained for a given density matrix $\varrho$. Baumgratz $et$ $al.$ \cite{baumgratz2014quantifying} has demonstrated that a quantum system's coherence is given as
\begin{equation}\label{coh03}
\mathcal{C}_{l_{1}}(\varrho)= \min_{\eta \in \mathfrak{I}} |\varrho - \eta|,
\end{equation}
where $\mathfrak{I}$ is a set containing all incoherent states, that is $\mathcal{C}_{l_{1}}(\eta) =0$ for all $\eta \in \mathfrak{I}$. The $l_{1}$-norm coherence can be obtained by means of the non-diagonal entries of $\varrho$ as
\begin{equation}
\mathcal{C}_{l_{1}}(\varrho) = \sum_{i\neq j} \sqrt{\varrho_{ij} \varrho_{ij}^*} \label{eq:norm-l1},
\end{equation}
where $^{*}$ refers to the complex conjugate. It should be noted that the total $l_1$-norm coherence of a state $\varrho$, living in  Hilbert space of dimension $d$, should not exceed  $d-1$. Correlated coherence is yet another metric of coherence that can provide information about the quantumness of a particular state. For any given bipartite state $\varrho$,  correlated coherence is defined as the total coherence subtracting local coherences. The definition of the $l_1$-norm based correlated coherence \cite{baumgratz2014quantifying} reads as
\begin{equation}
\mathcal{C}_{cc} (\varrho) := \mathcal{C}_{l_1} (\varrho)-\mathcal{C}_{l_1} (\varrho_A)-\mathcal{C}_{l_1} (\varrho_B),
\end{equation}
where $\varrho_A = tr_B \varrho$ and  $\varrho _B = tr_A \varrho$ are the reduced density matrices of local subsystems.

\subsection{Quantum discord}
QD \cite{ollivier2001quantum} quantifies quantum correlations inhibited in a composite system. In a two-qubit system, QD is defined by removing  the existing classical correlations from the quantum mutual information of the system  as
\begin{equation}
\label{eq17}
{\mathcal D}(\varrho)= {\mathcal I}(\varrho)-{\mathcal C}(\varrho),
\end{equation}
with
\begin{equation}
\label{eq18}
{\mathcal I}(\varrho)={\mathcal S}(\varrho_B)+ {\mathcal S}(\varrho_A)-{\mathcal S}(\varrho)
\end{equation}
and
\begin{equation}
\label{eq19}
{\mathcal C}(\varrho)={\mathcal S}(\varrho_A)- min_{\{\pi_{i}^{B}\}}\sum_{i}p_{i}{\mathcal S}(\varrho_{A|i}).
\end{equation}
$ {\mathcal S}(\varrho)=-tr(\varrho \log_2 \varrho) $ is the von Neumann entropy. $\varrho_A$ and $\varrho_B $ are the reduced density operators corresponding, respectively, to the subsystems $A$ and $B$.  $\/{\left\{ \pi_{B}^{i}\right\}}={|i_{B} \rangle \langle i_{B} |} $  is the complete ensemble  of orthonormal projectors acting only on the second subsystem $B$.\\
 $\varrho_{A|i} = tr_{B}(\pi_{B}^{i}\varrho \pi_{B}^{i}) / p_{i}$ is the resulting state of the first subsystem $A$ after obtaining the result $i$ on B, and $p_{i}= tr_{AB} ( \pi_B^i \varrho \pi_B^i)  $  is the probability of having $i$ as a result. Following the relations \eqref{eq18}  and \eqref{eq19}, the quantum discord \eqref{eq17} can be rewritten as \cite{henderson2001classical, fanchini2010non}
\begin{equation}
\displaystyle
 \label{eq2021}
{\mathcal D}(\varrho) = min_{               \left\lbrace \pi_B^i      \right\rbrace        }[{\mathcal S}(\varrho/{\left\{ \pi_{B}^{i}\right\}} ]+{\mathcal S}(\varrho_B)  -{\mathcal S}(\varrho).
 \end{equation}
In general, for arbitrary quantum states, finding analytical formulas for QD is complicated due to the minimization process required for conditional entropy. It was only possible to obtain approximate analytical expressions \cite{huang2013quantum} for a limited number of states, such as the X-states.  For $X$-states
\begin{equation}\label{eq01}
\varrho = {\begin{bmatrix}
{{\varrho _{11}}}&0&0&{{\varrho^* _{14}}}\\
0&{{\varrho _{22}}}&{{\varrho^* _{23}}}&0\\
0&{\varrho _{23}}&{{\varrho _{33}}}&0\\
{\varrho _{14}}&0&0&{{\varrho _{44}}}
\end{bmatrix}}
\end{equation}
the quantum discord is redefined by the succeeding expression \cite{wang2010classical, ali2010erratum}
\begin{equation}
\label{eq22}
{\mathcal D}(\varrho) = \min\{\mathcal{QD}_1, \mathcal{QD}_2\},
\end{equation}
with
$$ \mathcal{QD}_{i}= f(\varrho_{11 }+\varrho_{33 }) +\sum_{k=1}^{4} \lambda_k \log_2(\lambda_k)+{\mathcal D}_{i}.$$

${\mathcal D}_1= f(\Lambda)$, ${\mathcal D}_2=-\sum_{n=1}^{4}\varrho_{nn} \log_2(\varrho_{nn}) - f(\varrho_{11 }+\varrho_{33})$, $\Lambda =\frac{1}{2} \Big(1 + \Big((1-2(\varrho_{33 }+\varrho_{44 }))^2+4 (|\varrho_{14}|+|\varrho_{23}|)^2\Big)^{1/2} \Big)$ and $ f(x)=-x \log_2(x)-(1-x) \log_2(1-x)$ is the binary Shannon entropy.  $\lambda_k$'s denote the eigenvalues of the density matrix $\varrho$.

\section{Two-qubit XXZ  model  \label{sec3}}
We consider the model as a two-qubit system operating spin polarization of two nearest spin-$1/2$ XXZ particles. The particles are exposed to the interplay of an external homogeneous magnetic field and the combination of DM and KSEA interactions along the $z$-axis. The associated Hamiltonian is solvable and is represented as \cite{oumennana2022quantum}
\begin{eqnarray}
\label{Hamsys}
\hat{H} &=& J_z \sigma_{1}^{z}\sigma_{2}^{z}+ J(\sigma_{1}^{x}\sigma_{2}^{x}+ \sigma_{1}^{y}\sigma_{2}^{y})+ D_z\left(\sigma_{1}^{x}\sigma_{2}^{y}-\sigma_{1}^{y}\sigma_{2}^{x}\right) \nonumber\\
&+& \Gamma_z \left(\sigma_{1}^{x}\sigma_{2}^{y}+\sigma_{1}^{y}\sigma_{2}^{x}\right)+ B (\sigma_1^z + \sigma_2^z),
\end{eqnarray}
where $\sigma_{\mu}^{i=x,y,z}$ $(\mu=1,2)$  are the typical Pauli matrices corresponding to the spin $\mu$, while $J_z$ and $J$  are real coupling coefficients denoting, respectively, the anisotropy coupling constant defining the symmetric spin-spin exchange interaction in the $z$-direction, and the interaction coupling constant. The $B$ is a parameter,  restricted to $B \geq 0$, that indicates the strength of the magnetic field. Besides, the  $\Gamma_z$ and $D_z$ parameters reflect the $z$-KSEA and $z$-DM interactions, which result in symmetric and anti-symmetric spin-orbit coupling contributions, respectively. We assume that the two-spin XXZ model behaves as in the antiferromagnetic case, $J > 0$ and $J_z > 0$. It is worth stating that all parameters are considered to be dimensionless. The Hamiltonian  (\ref{Hamsys})  can be written on the two-qubit basis, $\lbrace \vert \Downarrow \Downarrow \rangle, \vert \Downarrow \Uparrow \rangle, \vert \Uparrow \Downarrow \rangle, \vert \Uparrow \Uparrow \rangle \rbrace$, as

\begin{equation}
\displaystyle
\label{Hmatrix}
   \hat{H} =
	 \left(
      \begin{array}{cccc}
      J_z+2B&0&0&-2i{\rm \Gamma}_z\\
      0&-J_z & 2 J+2iD_z&0\\
      0&-2iD_z + 2J &-J_z&0\\
      2i{\rm \Gamma}_z&0&0&-2B+J_z
      \end{array}
   \right).
\end{equation}

The eigenvalues and the associated eigenvectors of the aforementioned Hamiltonian $\hat{H}$ are
\begin{eqnarray*}
\mathcal{V}_1 &= J_z+ \chi, \quad \mathcal{V}_2 = -J_z+ \omega,\\
 & \mathcal{V}_3= -J_z- \omega,  \quad \mathcal{V}_4 = J_z- \chi,
\end{eqnarray*}

\begin{eqnarray*}
\vert u_1\rangle &= \sqrt{\frac{\chi+2B}{2\chi}}\left(|\Downarrow \Downarrow\rangle +\frac{ 2i{\rm \Gamma}_z}{{\chi+2B}}\vert\Uparrow \Uparrow\rangle\right),
\\
\vert u_2\rangle &= \frac{1}{\sqrt{2}}\left(|\Downarrow \Uparrow\rangle +\frac{2 J-2iD_z}{\omega} \vert\Uparrow \Downarrow\rangle\right),
\\
\vert u_3\rangle &= \frac{1}{\sqrt{2}} \left(|\Downarrow \Uparrow\rangle -\frac{2 J-2iD_z}{\omega}\vert\Uparrow \Downarrow\rangle\right),
\\
\vert u_4\rangle &= \sqrt{\frac{\chi-2B}{2\chi}}\left(|\Downarrow \Downarrow\rangle -\frac{2i{\rm \Gamma}_z}{\chi-2B}\vert\Uparrow \Uparrow\rangle\right),
\end{eqnarray*}
with $\chi = 2(B^2+\Gamma_{z}^{2})^{1/2}$ and $\omega=2(J^2+D_{z}^{2})^{1/2}$. To introduce intrinsic decoherence, we use Milburn's decoherence model  \cite{milburn1991intrinsic}, which assumes that quantum systems evolve continually in an arbitrary sequence of identical unitary transformations instead of unitary evolution. The following equation describes such evolution \cite{milburn1991intrinsic}
\begin{equation}
\frac{d\varrho_t}{dt} = \frac{1}{\gamma}\Bigl(\exp(-i\gamma \hat{H})\varrho_t \exp(i\gamma \hat{H})-\varrho_t\Bigr),
\label{milburn}
\end{equation}
here, $\varrho_t$ is the density matrix associated to the Hamiltonian $\hat{H}$, and $\gamma$ is the intrinsic decoherence constant. Thus in the limit of $\gamma^{-1} \rightarrow \infty$, there is no intrinsic decoherence, and Eq. \eqref{milburn} is reduced to the typical von Neumann equation characterizing an isolated quantum system. Milburn altered the Schrödinger equation in order for quantum coherence to be spontaneously destroyed throughout the evolution of the quantum system. The ensuing equation is obtained
\begin{equation}
\frac{d\varrho_t}{dt}=-\frac{\gamma}{2}[\hat{H},[\hat{H},\varrho_t]]-i [\hat{H},\varrho_t].
\label{milburn1}
\end{equation}
where $\frac{\gamma}{2}[\hat{H},[\hat{H},\varrho_t]]$ designates the non-unitary evolution under the intrinsic decoherence in our considered two-qubit system. A proper solution for the equation \eqref{milburn1} is obtained using the Kraus operators $\mathcal{\hat{M}}_{l}$ \cite{milburn1991intrinsic}

\begin{equation}
\varrho_t= \sum^{\infty}_{l=0}  \mathcal{\hat{M}}_l(t) \varrho^{t=0}\mathcal{\hat{M}}^{\dagger} _l{}(t)
\end{equation}
with $\varrho^{t=0}$ being the density matrix at $t=0$ of the considered system and $\mathcal{\hat{M}}_{l}$ are given by
\begin{align*}
\displaystyle
\mathcal{\hat{M}}_l(t) = \Big(\frac{\gamma^{l} t^{l}}{l!}\Big)^{1/2} { \hat{H}}^l \exp\left(-i{ \hat{H}}t\right) \exp\left(-\frac{\gamma t}{2}{\hat{H}}^{2}\right),
\end{align*}
with  $\sum^{\infty}_{l=0}  \mathcal{\hat{M}}_l(t) \mathcal{\hat{M}}^{\dagger}_l{}(t) =\mathbb{I}$. Finally, the evolved state of the two-spin XXZ quantum system described by $\hat{H}$ under intrinsic decoherence effects can be obtained by \cite{milburn1991intrinsic}

\begin{eqnarray}
\displaystyle
\varrho_t &=& \sum_{j,k} \exp\left(-\frac{\gamma t}{2}({\mathcal{V}}_j-{\mathcal{V}}_{k})^{2}- i({\mathcal{V}}_{j}-{\mathcal{V}}_{k})t\right) \nonumber\\
&\times& \langle u_{j}\vert\varrho^{t=0}\vert u_{k}\rangle \vert u_{j}\rangle\langle u_{k}\vert,
\label{sol}
\end{eqnarray}
where ${\mathcal{V}}_{j,k}$ and $\ket{u_{j,k}}$ are, respectively, the eigenvalues of the Hamiltonian $\hat{H}$ (\ref{Hmatrix}) and their corresponding eigenstates. Eq. (\ref{sol}) allows us to describe how the system resources change as the state of the system is evolving under the intrinsic decoherence effects. Following, we use correlated coherence and quantum discord to study how intrinsic decoherence affects coherence and nonclassical correlations in the two-spin XXZ Heisenberg model under the different supposed interactions. For a complete characterization of the influence of intrinsic decoherence on the system, we take into account two possible scenarios of the initial state
\begin{eqnarray}
\label{bell1}
\ket{\Psi_1} &=& \cos\left(\theta/2\right)\ket{\Downarrow \Downarrow}+\sin\left(\theta/2\right) \ket{\Uparrow \Uparrow}, \\
\ket{\Psi_2} &=& \cos\left(\theta/2\right)\ket{\Downarrow \Uparrow}+\sin\left(\theta/2\right) \ket{\Uparrow \Downarrow}.
\label{bell2}
\end{eqnarray}
where $\ket{0} \equiv \ket{\Downarrow}$ and $\ket{1} \equiv \ket{\Uparrow}$, denoting spin-down and spin-up, respectively. We shall study intrinsic decoherence according to two samples of Bell-like initial states. To do that, we combine the states in Eq. \eqref{bell1}-\eqref{bell2} using a parametrization and then solve the dynamical equation. The results for each sample initial state can be obtained by dealing with the parametrization parameters.  Thus, the initial state looks as follows
\begin{equation}\label{wer}
\varrho^{t=0} = \alpha_1 \varrho^{t=0} _1 + \alpha_2 \varrho^{t=0} _2,
\end{equation}
where $\varrho^{t=0} _i = p \ket{\Psi_i}\bra{\Psi_i} + (1-p)/4 ~ I_4$, $0 \leq p \leq 1$ denoting the level of purity in the initial state, $0 \leq \theta <\pi$ and $\alpha_1$ and $\alpha_2$ are the parametrization parameters taking either zero or one in each of the two scenarios. We note that when Bell-like states reduce to Bell states, those of EWL states reduce to Werner states \cite{werner1989quantum}. The parameterized initial state is as follows
 \begin{widetext}
\begin{align*}
\displaystyle
\label{ewl01}
\varrho^{t=0} = \left(
\begin{array}{cccc}
 \frac{1}{4} \left(4 \alpha _1^2 p \cos ^2\left(\frac{\theta }{2}\right)-p+1\right) & 0 & 0 & \alpha _1^2 p \sin
   \left(\frac{\theta }{2}\right) \cos \left(\frac{\theta }{2}\right) \\
 0 & \frac{1}{4} \left(4 \alpha
   _2^2 p \cos ^2\left(\frac{\theta }{2}\right)-p+1\right) & \alpha _2^2 p \sin
   \left(\frac{\theta }{2}\right) \cos \left(\frac{\theta }{2}\right) & 0 \\ 0 & \alpha _2^2 p \sin \left(\frac{\theta }{2}\right) \cos \left(\frac{\theta
   }{2}\right) & \frac{1}{4} \left(4 \alpha _2^2 p \sin ^2\left(\frac{\theta
   }{2}\right)-p+1\right) & 0 \\
 \alpha _1^2 p \sin \left(\frac{\theta }{2}\right) \cos \left(\frac{\theta }{2}\right) &
   0 & 0 & \frac{1}{4}
   \left(4 \alpha _1^2 p \sin ^2\left(\frac{\theta }{2}\right)-p+1\right)
\end{array}
\right).
\end{align*}
 \end{widetext}
To simplify calculations, all terms that correspond to the overlapping $\alpha_1 \alpha_2$ in $\varrho^{t=0}$ cancel in both situations and are thus omitted. By solving Milburn's equation, one finds that the time-dependent density matrix encoding both two cases takes the form
\begin{equation}
\label{evolved}
\varrho^t = \begin{pmatrix}
	\varrho_{11} & 0 & 0 & \varrho_{14} \\
	0 & \varrho_{22} & \varrho_{23} & 0 \\
	0 & \varrho_{23}^{\ast} & \varrho_{33} & 0 \\
	\varrho_{14}^* & 0 & 0 & \varrho_{44}
\end{pmatrix}.
\end{equation}
The time-evolution preserves the X-structure of the initial extended Werner-like (EWL) state for any choice between the two initial states.  The following table gives the properties of the density matrix for each case.
\begin{table}[H]
{\begin{tabular}{@{}cccc@{}} \toprule
$\ket{\Psi_i}$ \hphantom{$\Downarrow \Downarrow$}& $(\alpha_1, \alpha_2)$ & $\varrho_{ij} \neq 0$ \\
\colrule
\\
$\ket{\Psi_1}$ \hphantom{$\Downarrow \Downarrow$} & $(1, 0)$  & ($\varrho_{11}, \varrho_{14}, \varrho_{22}, \varrho_{44})$, with $\varrho_{33} = \varrho_{22}, \varrho_{41} = \varrho_{14}^*$
\\ \\ \botrule
$\ket{\Psi_2}$ \hphantom{$\Downarrow \Downarrow$} & $(0, 1)$  & ($\varrho_{11}, \varrho_{22}, \varrho_{23}, \varrho_{33})$, with $\varrho_{11} = \varrho_{44}, \varrho_{32} = \varrho_{23}^*$ \\ \\ \botrule
\end{tabular}\label{ta1}}
\end{table}
For the combined initial state in \eqref{wer}, we get the following density matrix entries
\begin{widetext}
\begin{subequations}\label{varrho-entries}
\begin{align*}
\displaystyle
\varrho_{11} &= \frac{\alpha _1}{\chi ^2} \left(B^2 (2 p \cos (\theta )+p+1)+p \Gamma _z e^{-2 \gamma  t \chi ^2} \left(2 \cos (\theta ) \Gamma _z \cos (2 t \chi )-\chi  \sin (\theta ) \sin (2 t \chi )\right)+(p+1) \Gamma _z^2\right)-\frac{1}{4} \alpha _2 (p-1),
\\
\varrho_{14} &= \frac{\alpha _1 p e^{-2 \gamma  t \chi ^2}}{2 \chi ^2} \left(\chi  \sin (\theta ) (\chi  \cos (2 t \chi )-2 i B \sin (2 t \chi ))+2 \cos (\theta ) \Gamma _z \left(-2 i B e^{2 \gamma t \chi ^2}+2 i B \cos (2 t \chi )+\chi  \sin (2 t \chi )\right)\right),
\\
\varrho_{22} &=  \frac{\alpha _2}{4 \omega} \Big(\left(2 p e^{-2 \gamma  t \omega ^2} \Big(\left(2 D_z \sin (\theta ) \sin (2 t \omega )+\omega  \cos (\theta ) \cos (2 t \omega
   )\right)+(p+1) \omega \right)-\alpha _1 (p-1) \omega \Big)\Big),
\\
\varrho_{23} &= \frac{\alpha _2 p  e^{-2 \gamma  t \omega ^2}}{\omega ^2} \left(J-i D_z\right) \left(-2 i D_z \sin(\theta ) \cos (2 t \omega )+2 J \sin (\theta ) e^{2 \gamma  t \omega ^2}+i \omega \cos (\theta ) \sin (2 t \omega )\right),
\\
\varrho_{33} &=  \frac{\alpha _2}{4 \omega} \Big( \left((p+1) \omega -2 p e^{-2 \gamma  t \omega ^2} \left(2 D_z \sin (\theta ) \sin (2 t \omega )+\omega  \cos (\theta ) \cos (2 t \omega )\right)\right)-\alpha _1 (p-1) \omega \Big),
\\
\varrho_{44} &= \frac{\alpha _1}{\chi^2} \left(B^2 (-2 p \cos (\theta )+p+1)+p \Gamma _z e^{-2 \gamma  t \chi ^2} \left(\chi  \sin (\theta ) \sin (2 t \chi )-2 \cos (\theta ) \Gamma _z \cos (2 t \chi )\right)+(p+1) \Gamma _z^2\right)-\frac{1}{4} \alpha _2 (p-1).
\end{align*}
\end{subequations}
\end{widetext}
The populations of the system combining both scenarios are given implicitly as
\begin{eqnarray*}
\displaystyle
\lambda_1 &=& \frac{1}{2} \left(-\sqrt{4 \left| \varrho _{23}\right| {}^2+\left(\varrho _{22}-\varrho _{33}\right){}^2}+\varrho _{22}+\varrho _{33}\right),\\
\lambda_2 &=& \frac{1}{2} \left(\sqrt{4 \left| \varrho _{23}\right| {}^2+\left(\varrho _{22}-\varrho _{33}\right){}^2}+\varrho _{22}+\varrho _{33}\right),
\end{eqnarray*}
\begin{eqnarray*}
\displaystyle
\lambda_3 &=&\frac{1}{2} \left(-\sqrt{4 \left| \varrho _{14}\right| {}^2+\left(\varrho _{11}-\varrho _{44}\right){}^2}+\varrho _{11}+\varrho _{44}\right),\\
\lambda_4 &=& \frac{1}{2} \left(\sqrt{4 \left| \varrho _{14}\right| {}^2+\left(\varrho _{11}-\varrho _{44}\right){}^2}+\varrho _{11}+\varrho _{44}\right).
\end{eqnarray*}
We use the following notation to locate the matrix density and populations for each case:
\begin{equation}
\label{rcase}
\varrho^t =
    \begin{cases}
      \varrho^t _1, & \text{if}\ \alpha_1 = 1, \alpha_2 = 0 \\
      \varrho^t _2, & \text{if}\ \alpha_1 = 0, \alpha_2 = 1
    \end{cases}.
 \end{equation}
By combining the two cases in \eqref{rcase}, correlated coherence writes as
\begin{eqnarray}
\displaystyle
\mathcal{C}_{cc} (\varrho^t)  &=& \frac{\alpha _1 p e^{-2 \gamma  t \chi ^2}}{\chi ^2} \Bigg[\Big(2 B \chi  \sin (\theta ) \sin (2 t \chi )+4 B \cos (\theta )\nonumber
\\
&\times& \Gamma _z \left(e^{2 \gamma  t \chi ^2}-\cos (2 t \chi )\right)\Big){}^2\nonumber
\\
&+&\left(\chi ^2 \sin (\theta ) \cos (2 t \chi )+2 \chi  \cos (\theta ) \Gamma _z \sin (2 t \chi )\right){}^2\Bigg]^{1/2}\nonumber
\\
&+&\frac{\alpha _2 p e^{-2 \gamma  t \omega ^2}}{\omega } \Bigg[\Big(\omega  \cos (\theta ) \sin (2 t \omega )-2 D_z \sin (\theta ) \nonumber
\\
&\times& \cos (2 t \omega )\Big){}^2+4 J^2 \sin ^2(\theta ) e^{4 \gamma  t \omega ^2}\Bigg]^{1/2}.
\end{eqnarray}
One can check that $\varrho_{14}$ and $\varrho_{23}$ are the sole non-zero off-diagonal elements, storing all information about the system's coherence for each studied case. As well, one can easily verify that local coherence, ($\mathcal{C}^L (\varrho) = \mathcal{C}_{l_1} (\varrho_A) + \mathcal{C}_{l_1} (\varrho_B) = 0$), is always zero and thus correlated coherence is the system's total coherence.  Similarly, quantum discord is obtained implicitly as
\begin{eqnarray}
\normalsize
\label{qd1} \mathcal{QD}_{1} &=& \log_2 \left(\frac{(1-\Lambda )^{\Lambda }}{\Lambda ^{\Lambda }-\Lambda ^{\Lambda +1}} \prod_{i=1}^4 \lambda _i^{\lambda _i} \left(\frac{\beta}{1-\beta}\right){}^{-\beta}\right)\nonumber\\
 &-& \log_2 \left(-\beta+1\right),
  \\
 \label{qd2} \mathcal{QD}_{2} &=&  \log_2 \left(\prod_{i=1}^4 \lambda _i^{\lambda _i} \times \prod_{j=1}^4 \varrho _{jj}^{-\varrho _{jj}}  \right),
 \end{eqnarray}
where $\Lambda = \frac{1}{2} \left(\sqrt{4 \left(\varrho _{14}+\varrho _{23}\right){}^2+\left(1-2 \left(\varrho _{33}+\varrho _{44}\right)\right){}^2}+1\right)$, $\beta = \varrho _{11}+\varrho _{33}$ and $\lambda_i$'s are the populations of the system. Quantum discord is then computed by plugging Eq. \eqref{qd1} and Eq. \eqref{qd2} in Eq. \eqref{eq22}.

\section{Results and analysis \label{sec4}}
This section presents the obtained findings regarding correlated coherence and quantum discord. To sweeten our learning of the dynamical aspect of quantum correlations and coherence in the two-qubit XXZ Heisenberg model, we considered two kinds of EWL states. We then depict the behavior of coherence and quantum correlations as a function of time and all the parameters in the Hamiltonian as well as the level of purity of the initial state, the Bloch angle, and the decoherence rate.

\begin{figure}[H]
\centering
\subfigure[]{\label{figure1a}\includegraphics[scale=0.6]{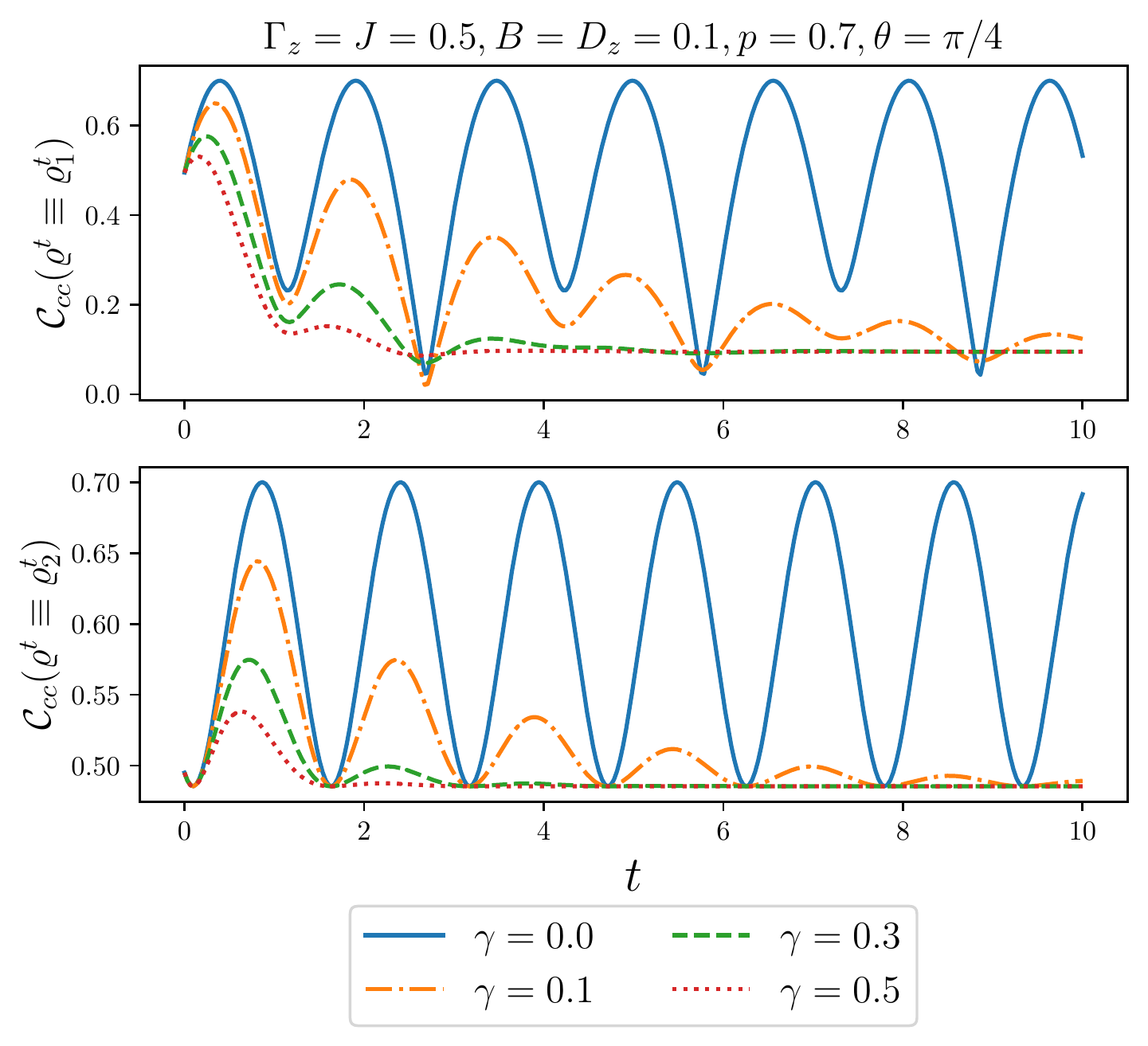}}
\subfigure[]{\label{figure1b}\includegraphics[scale=0.6]{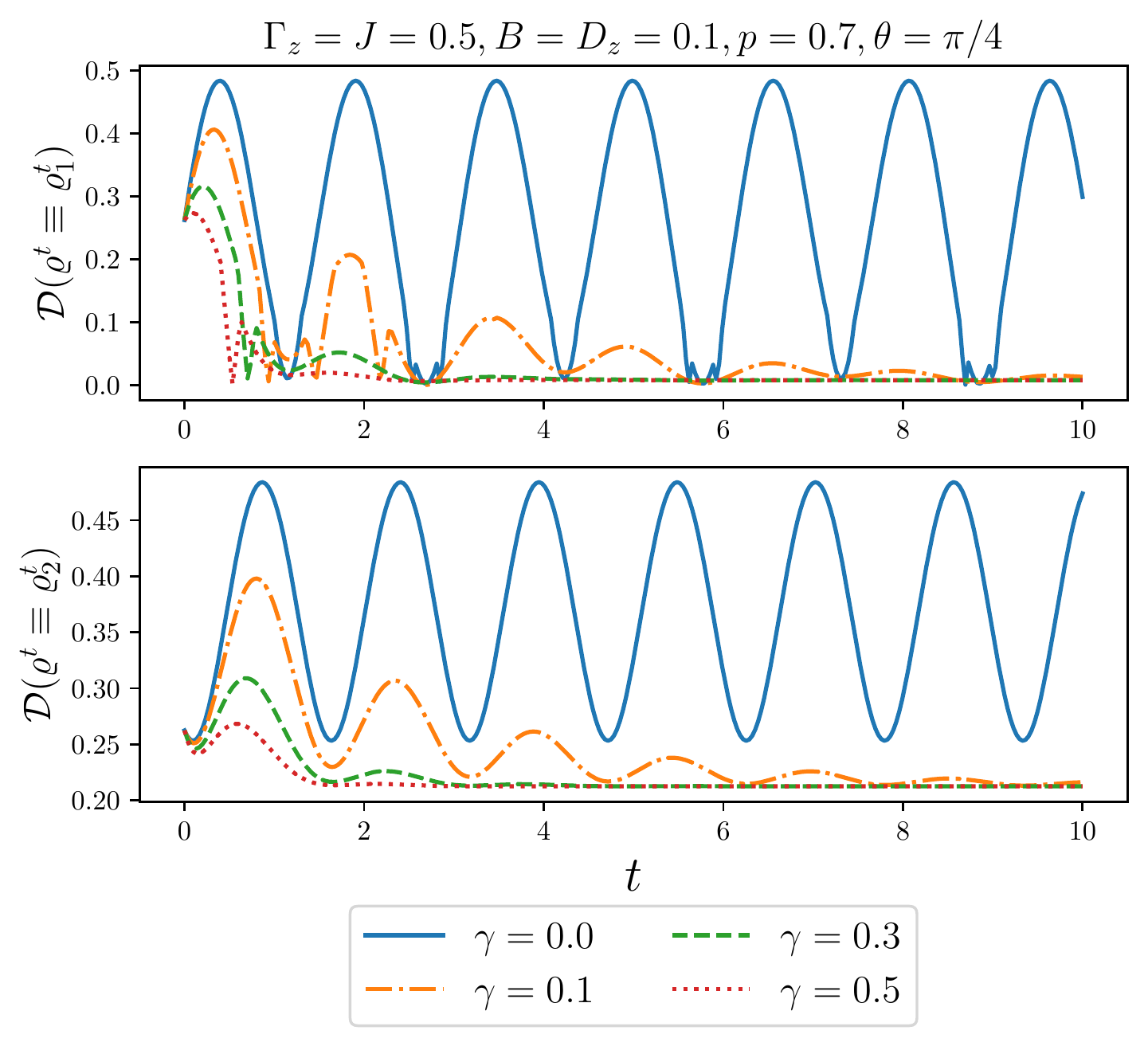}}
\caption{Time-evolution of coherence \textit{(correlated coherence)} \ref{figure1a} and quantum correlations \textit{(quantum discord)} \ref{figure1b}. Case 1 ($\varrho^t \equiv \varrho^t _1$) and case 2 ($\varrho^t \equiv \varrho^t _2$), versus decoherence rates $\gamma$.}
\label{figure1}
\end{figure}
In Fig. \ref{figure1} we plot the correlated coherence and quantum discord as functions of time $t$ for both initial EWL states $\varrho_{1}^{t=0}$ and $\varrho_{2}^{t=0}$ for some decoherence rates $\gamma$.  We notice that both measures show oscillatory behavior due to intrinsic decoherence. At the initial time $t=0$, the two quantifiers start from the same equal nonzero value for both initial states $\varrho_{1}^{t=0}$ and $\varrho_{2}^{t=0}$, that is $\mathcal{C}_{cc}(\varrho_{1}^{t=0}) = \mathcal{C}_{cc}(\varrho_{2}^{t=0}) \approx 0.495$ and $\mathcal{D}(\varrho_{1}^{t=0}) = \mathcal{D}(\varrho_{2}^{t=0}) \approx 0.262$. Without intrinsic decoherence $\gamma=0$, both measures display non-damping oscillations and fluctuates between their minimum and maximum values for an infinite time scale. For $\gamma > 0$, as time increases, correlated coherence and quantum discord experience damped oscillations and get closer to a stable value after an adequate period suggesting that the system has evolved to a steady state.

The attained steady state is not dependent on $\gamma$ but on the initial Werner-like state and the Hamiltonian parameters, namely $p$ and $\theta$. Also, it is marked that a more considerable $\gamma$ drives the quick decline of correlated coherence and discord; it causes the number and size of their oscillations to drop at some point, causing the system's state to quickly transition into a steady state. The coherence and correlation deterioration are due to the damping terms $e^{-2 \gamma  t \chi ^2}$  (in the case of $\varrho_{1}^t$) and $e^{-2 \gamma  t \omega ^2}$ (in the case of $\varrho_{2}^t$); these functions reduce the coherence and quantum correlations amplitude after each incrementing $t$ or $\gamma$. Similarly, we can mark identical routines for all the different cases.

In the asymptotic limit $t \to \infty$, quantum coherence and quantum correlations reach their stable value. The higher the steady-state correlation, the more the system will be appropriate for achieving quantum-based tasks. In the same conditions of $\Gamma_z = J = 0.5, B = D_z = 0.1, p = 0.7, \theta = \pi/4$ and $\gamma > 0$, we numerically find the value of correlated coherence as $\mathcal{C}_{cc}(\varrho_{1}^{t \to \infty}) \approx 0.095$ and $\mathcal{C}_{cc}(\varrho_{2}^{t \to \infty}) \approx 0.485$. Similarly, for quantum discord, we find $\mathcal{D}(\varrho_{1}^{t \to \infty}) \approx 0.007$ and $\mathcal{D}(\varrho_{2}^{t \to \infty}) \approx 0.212$.

Remarkably, we encountered that $\mathcal{C}_{cc}(\varrho_{1}^{t \to \infty})<\mathcal{C}_{cc}(\varrho_{2}^{t \to \infty})$  and $\mathcal{D}(\varrho_{1}^{t \to \infty}) <  \mathcal{D}(\varrho_{2}^{t \to \infty})$, this means that the system sustains more superposition and correlations when it is prepared in the initial state $\varrho_{2}^{t = 0}$. On the other hand, when the system is initially prepared in the Werner-like state $\varrho_{1}^{t = 0}$, its steady state is still a weakly correlated mixed state that sustains low superposition.

\begin{widetext}
\begin{minipage}{\linewidth}
\begin{figure}[H]
\centering
\subfigure[]{\label{figure2a}\includegraphics[scale=0.6]{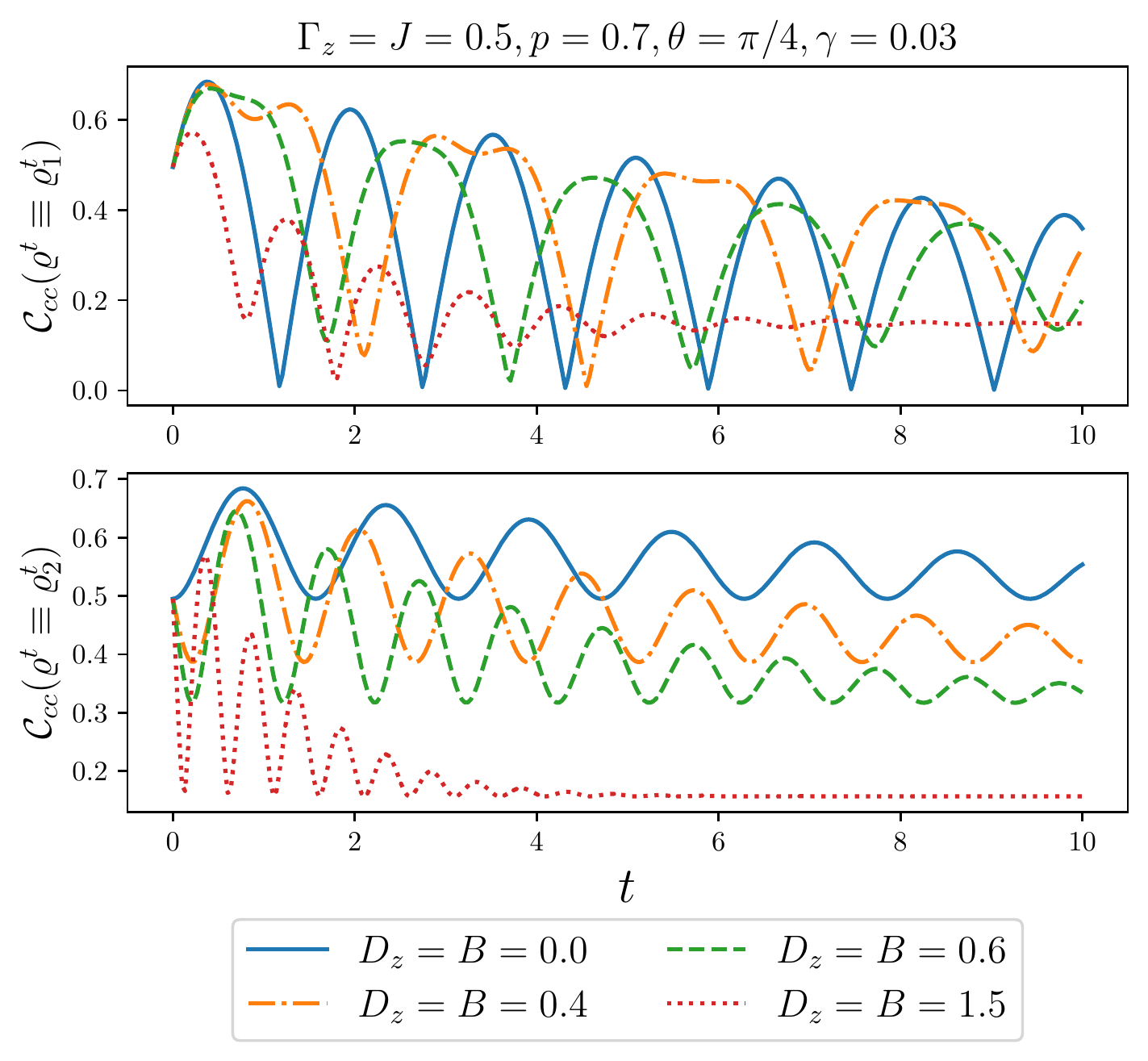}}
\subfigure[]{\label{figure2b}\includegraphics[scale=0.6]{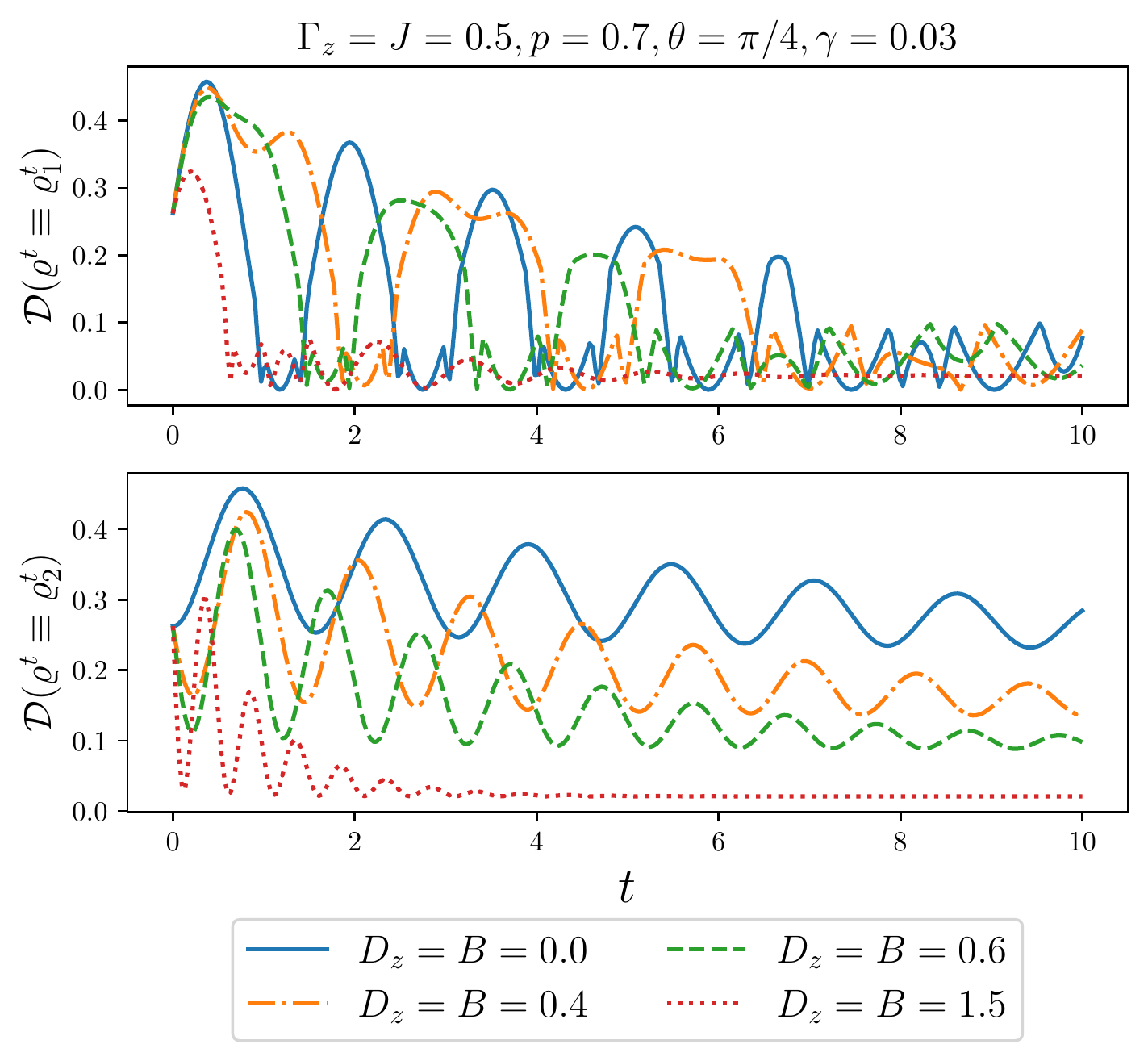}}
\caption{Time-evolution of coherence \textit{(correlated coherence)} \ref{figure2a} and quantum correlations \textit{(quantum discord)} \ref{figure2b}. Case 1 ($\varrho^t \equiv \varrho^t _1$): versus $B$. Case 2 ($\varrho^t \equiv \varrho^t _2$): versus $D_z$.}
\label{figure2}
\end{figure}
\end{minipage}
\end{widetext}
Fig.\ref{figure2} shows the dynamics of correlated coherence and quantum discord versus time $t$ for various values of the DM interaction strength $D_{z}$ and the homogeneous magnetic field $B$. It is worth noting, as it was shown analytically above that the DM interaction does not affect the system when the considered initial state is $\varrho_{1}^{t=0}$ as $D_{z}$ does not appear in the expressions of the entries of the evolved density matrix $\varrho_{1}^{t}$. Likewise, the effect of the magnetic field $B$ can not be studied for $\varrho_{2}^{t}$ since the structure of the initial EWL state $\varrho_{2}^{t=0}$ totally suppresses its contribution. 

In Fig. \ref{figure2a}, we notice that for growing intensities of the magnetic field $B$, the oscillatory behavior of correlated coherence within $\varrho_{1}^{t}$ swiftly declines as the amplitudes are shrinking, and $\mathcal{C}_{cc} (\varrho_{1}^{t})$ rapidly stabilizes on a nonzero frozen state. Furthermore, the oscillations are slightly phase-shifted, and when $B\neq 0$, the peaks presenting the maximal values reached over time are broader for more diminutive intensities of $B$. We again observe that the lower bound of the oscillations is unsteady as it can increase or decrease over time. When the external magnetic field is non-existing ($B=0$), correlations are damped over time due to intrinsic decoherence, and we regard the collapse and regeneration phenomena. On the other hand, quantum discord (Fig.  \ref{figure2b}) exhibits erratic oscillations; they are indeed damped over time but present successive small and irregular spikes. Physically, since the Milburn model is characterized by the associated magnetic field $B$, the irregular oscillatory behavior of correlated coherence and discord in the case $\varrho ^t \equiv \varrho _1 ^t$, could be due to an ongoing information flow between the magnetic field and the two-qubit XXZ spin system. For the second density matrix $\varrho^{t}\equiv \varrho_{2}^{t}$, we see that the same behavior is manifested by both quantum measures with the mere difference that oscillations are regular and that correlated coherence records higher values than quantum discord. At $t=0$, the quantum correlations and correlated coherence amounts are nonzero. Their initial recorded quantities do not depend on the DM interaction's strength $D_{z}$ since they are fixed when the initial state is given. In the absence of the DM interaction $D_{z}=0$, both measures record the highest values. In contrast, for increasing $D_{z}$ values, the oscillations' frequency increases despite being quickly damped. The oscillations get damped and saturate to a stable value for a finite time value. This might be explained through the many-body localization-delocalization viewpoint \cite{doggen2018many}.  Meanwhile, for $\varrho^{t}\equiv \varrho_{1}^{t}$ and $\gamma>0$, the steady states of correlated coherence and discord are closely related to both $B$ and $\Gamma_z$, it turns out that there is a competition between $B$ and $\Gamma_z$. In more specific terms, the steady states increase/decrease by increasing/decreasing either $B$ or $\Gamma_z$ until $B = \Gamma_z$, the point at which the steady state reaches its maximum value. This result indicates that $B$ and $\Gamma_z$ cancel each other so that the steady states of correlated coherence and discord remain dependent only on $p$ and $\theta$. We note that for all $\gamma > 0$, the maximum steady state recorded for $\theta = \pi/4$ and $p = 0.7$ remains constant for all nonzero intensities $B = \Gamma_{z}$, numerically, $\mathcal{C}_{cc} (\varrho_1 ^{t \to \infty}) \approx 0.2474$ and $\mathcal{D} (\varrho_1 ^{t \to \infty}) \approx 0.0543$.

For the second configuration $\varrho^{t}\equiv \varrho_{2}^{t}$, we see that as the DM interaction strength grows, the value of the steady state decreases. In contrast, as the spin-spin interaction coupling becomes strong, the steady state of correlated coherence and quantum discord improves.  For $\gamma >0$ and $J=D_z \neq 0$, the steady state can be modified by changing $p$ and $\theta$. These results highlight the impact of the initial state on the whole evolution of the system and suggest that its quantum resources could be tweaked  and enhanced through the modification of certain parameters.
\begin{widetext}
\begin{minipage}{\linewidth}
\begin{figure}[H]
\centering
\subfigure[]{\label{figure3a}\includegraphics[scale=0.6]{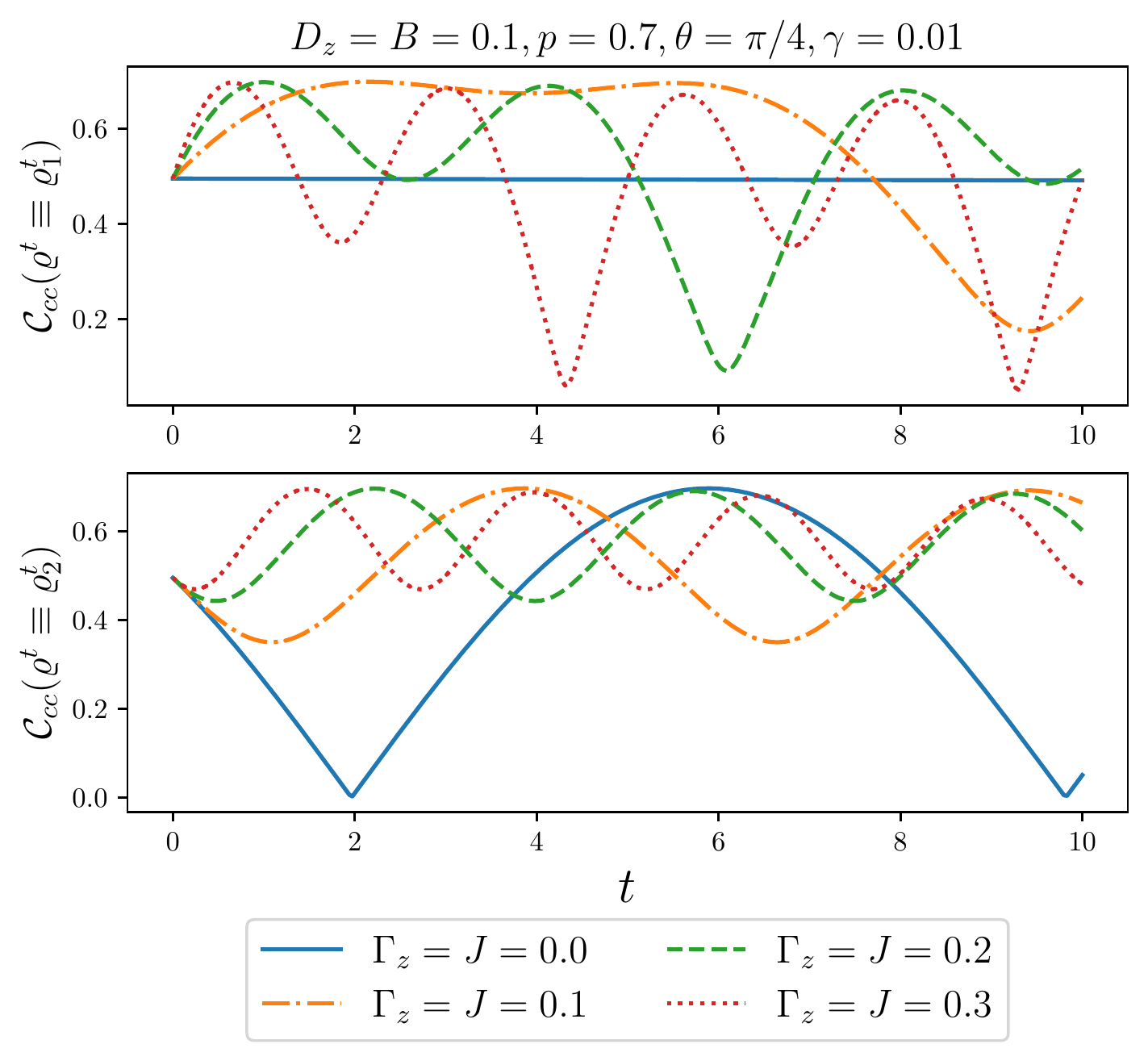}}
\subfigure[]{\label{figure3b}\includegraphics[scale=0.6]{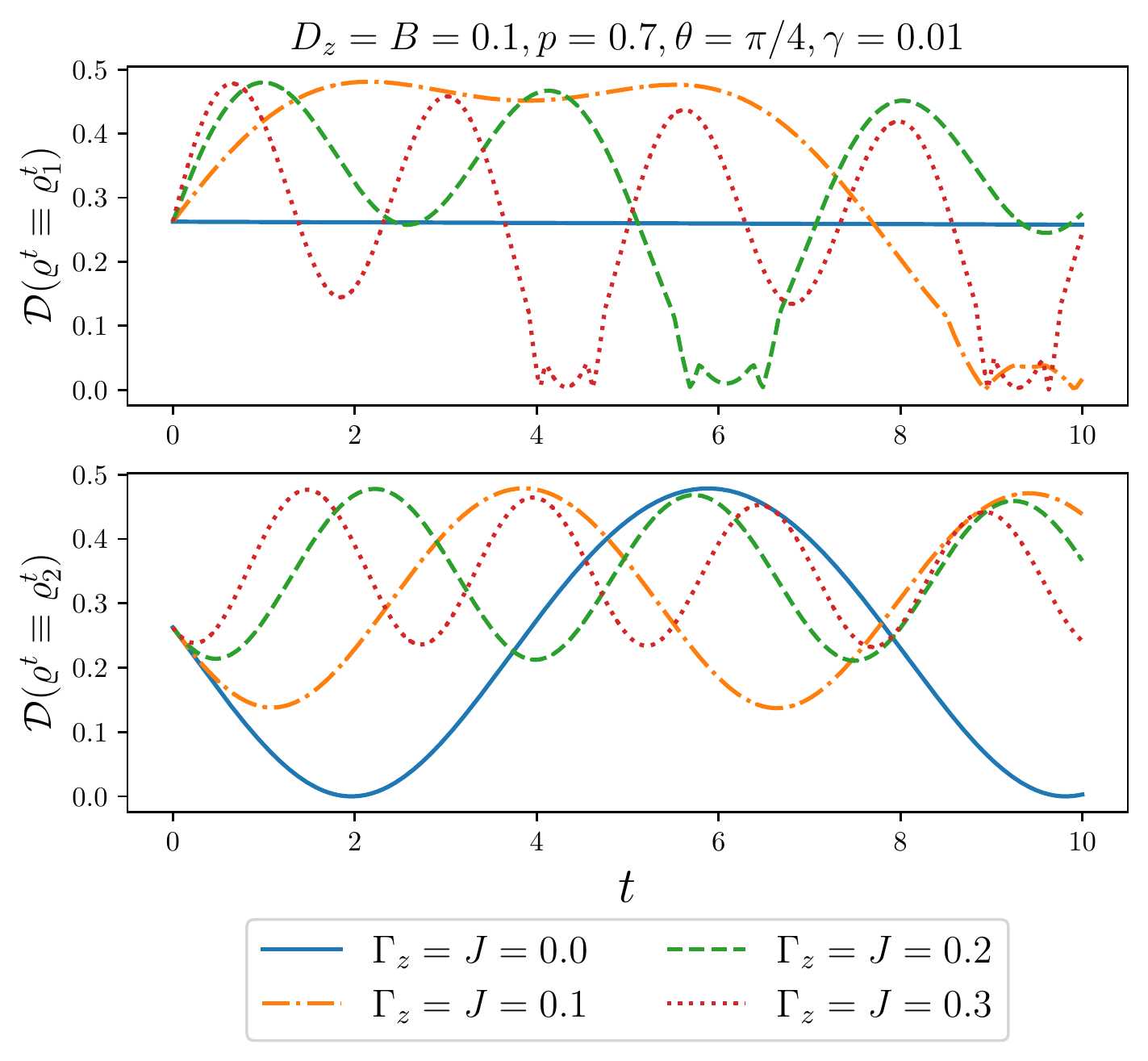}}
\caption{Time-evolution of coherence \ref{figure3a} and quantum correlations \ref{figure3b}. Case 1 ($\varrho^t \equiv \varrho^t _1$): versus $\Gamma _z$. Case 2 ($\varrho^t \equiv \varrho^t _2$): versus $J$.}
\label{figure3}
\end{figure}
\end{minipage}
\end{widetext}
In Fig. \ref{figure3}, we depict the influence of the strength of the KSEA interaction ($\Gamma_{z}$) and the spin-spin coupling $J$ on correlated coherence and quantum discord for both considered initial states. We specify again that the effect of the first parameter is examined only in the first case where $\varrho^{t}\equiv \varrho_{1}^{t}$, while the influence of the anisotropy parameter $J$ can solely be examined when $\varrho^{t}\equiv \varrho_{2}^{t}$.

The first salient observation regarding the evolution of correlated coherence and quantum discord in the first considered case $\varrho^{t}\equiv \varrho_{1}^{t}$, is that both appear to be stable for $\Gamma_{z}=0$ for short periods. However, they eventually decrease for sufficiently long periods. In the absence of the KSEA interaction, the expression of correlated coherence reduces to $\mathcal{C}_{cc} (\varrho_{1}^{t})=pe^{-2\gamma t\times 4B^{2}}\sin{\theta}$, and we can numerically perceive its time evolution. Correlated coherence exhibits the same oscillatory pattern in the presence of the KSEA interaction ($\Gamma_{z}\neq0$). However, as $\Gamma_{z}$ decreases, the lower bound of oscillations is higher, and the minimum values shift to the right. As KSEA interaction strengthens, the frequency of oscillations decreases, and coherence becomes less erratic. Quantum discord displays identical behavior as correlated coherence with lower quantities. In distinction, we spot in the dynamical behavior of QD the resurgence of the collapse and revival phenomena and the previously mentioned pattern of minor successive increases and decreases when oscillations reach their minimum values.

In the second case $\varrho^{t}\equiv \varrho_{2}^{t}$, we observe that for increasing interaction coupling constant $J$, the frequency of oscillations of both quantifiers increases, but the amplitudes shrink. For $J=0$, the lower bounds of both measures nearly drop to zero at the same instants $t$, at $t\approx2$ and $t\approx10$, and we see that correlated coherence displays sharp peaks at these instants. Once again, it is apparent that the initial amounts of quantum correlations and correlated coherence are not related to the parameters that do not figure in the initial state.

\begin{widetext}
\begin{minipage}{\linewidth}
\begin{figure}[H]
\centering
\subfigure[]{\label{figure4a}\includegraphics[scale=0.6]{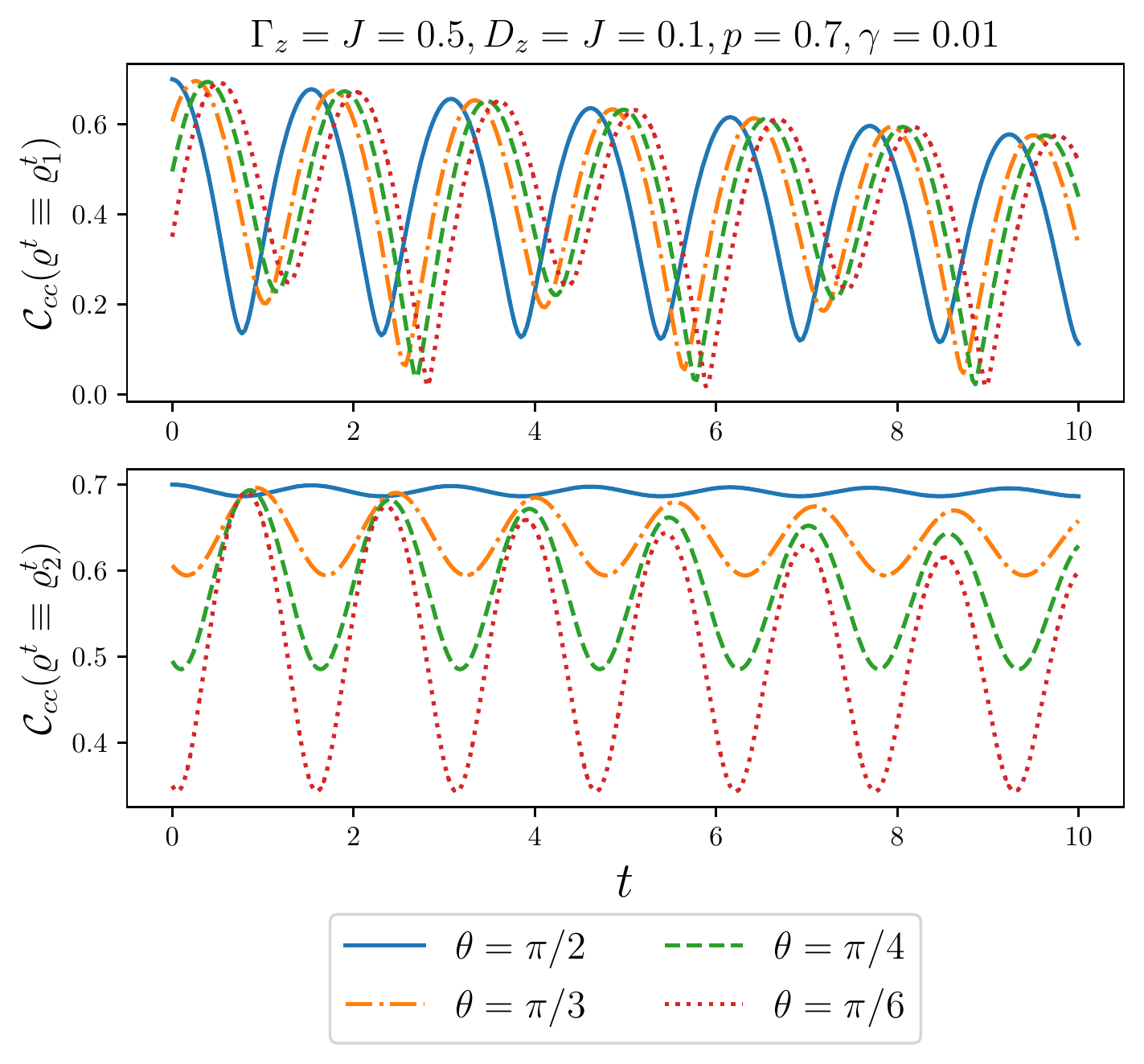}}
\subfigure[]{\label{figure4b}\includegraphics[scale=0.6]{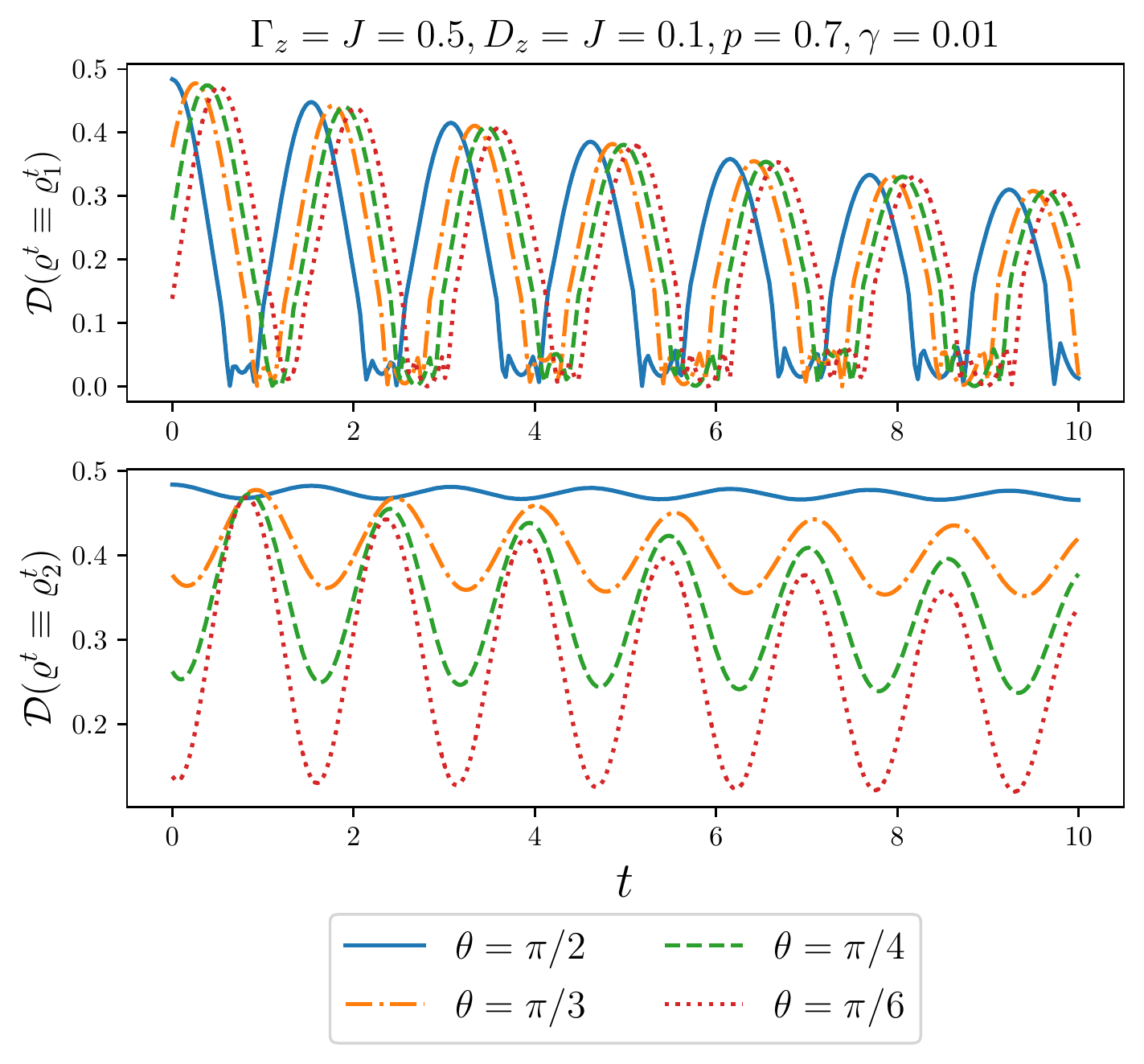}}
\caption{Time-evolution of coherence \ref{figure4a} and quantum correlations \ref{figure4b}. Case 1 ($\varrho^t \equiv \varrho^t _1$) and case 2 ($\varrho^t \equiv \varrho^t _2$), versus the Bloch angle $\theta$.}
\label{figure4}
\end{figure}
\end{minipage}
\end{widetext}
In Fig. \ref{figure4}, we visualize the effect of the angle $\theta$ on correlated coherence and quantum discord dynamics. Unlike the previous figures where the initial values, at $t=0$, of $\mathcal{C}_{cc} (\varrho^{t})$ and ${\mathcal D}(\varrho^{t})$ are not dependent on the strengths of the parameters $\gamma$ (Fig. \ref{figure1}), $D_{z}$ and $B$ (Fig. \ref{figure2}), $J$ and $\Gamma_{z}$  (Fig. \ref{figure3}), we notice in Fig. \ref{figure4} that the initial amounts of quantum correlations and correlated coherence are related to the value of the angle $\theta$.

Clearly, as the value of $\theta$ rises, not only the initial recorded values are higher, but also those recorded during the time evolution. The angle $\theta$ does not affect the frequency of oscillations, but it  affects the quantities of correlated coherence and nonclassical correlations within the system. Moreover, oscillations are in phase when $\varrho^{t}\equiv \varrho_{2}^{t}$, unlike the first case $\varrho^{t}\equiv \varrho_{1}^{t}$ where oscillations are slightly phase-shifted to the right for decreasing $\theta$ values.

For $\theta=\frac{\pi}{2}$, which is the angle value for which the Bell-like states (\ref{bell1}-\ref{bell2}) reduce to the maximally entangled Bell states and the EWL becomes a Werner state, $\mathcal{C}_{cc} (\varrho^{t})$ and ${\mathcal D}(\varrho^{t})$, present the higher quantities but they are roughly invariant as they present timid fluctuations, when $\varrho^{t}\equiv \varrho_{2}^{t}$. Another worth noting remark is the collapse and revival phenomena exhibited in the dynamical behavior of quantum discord when $\varrho^{t}\equiv \varrho_{1}^{t}$, in addition to the pattern of the small spikes in the lower bound of QD.
\begin{widetext}
\begin{minipage}{\linewidth}
\begin{figure}[H]
\centering
\subfigure[]{\label{figure5a}\includegraphics[scale=0.6]{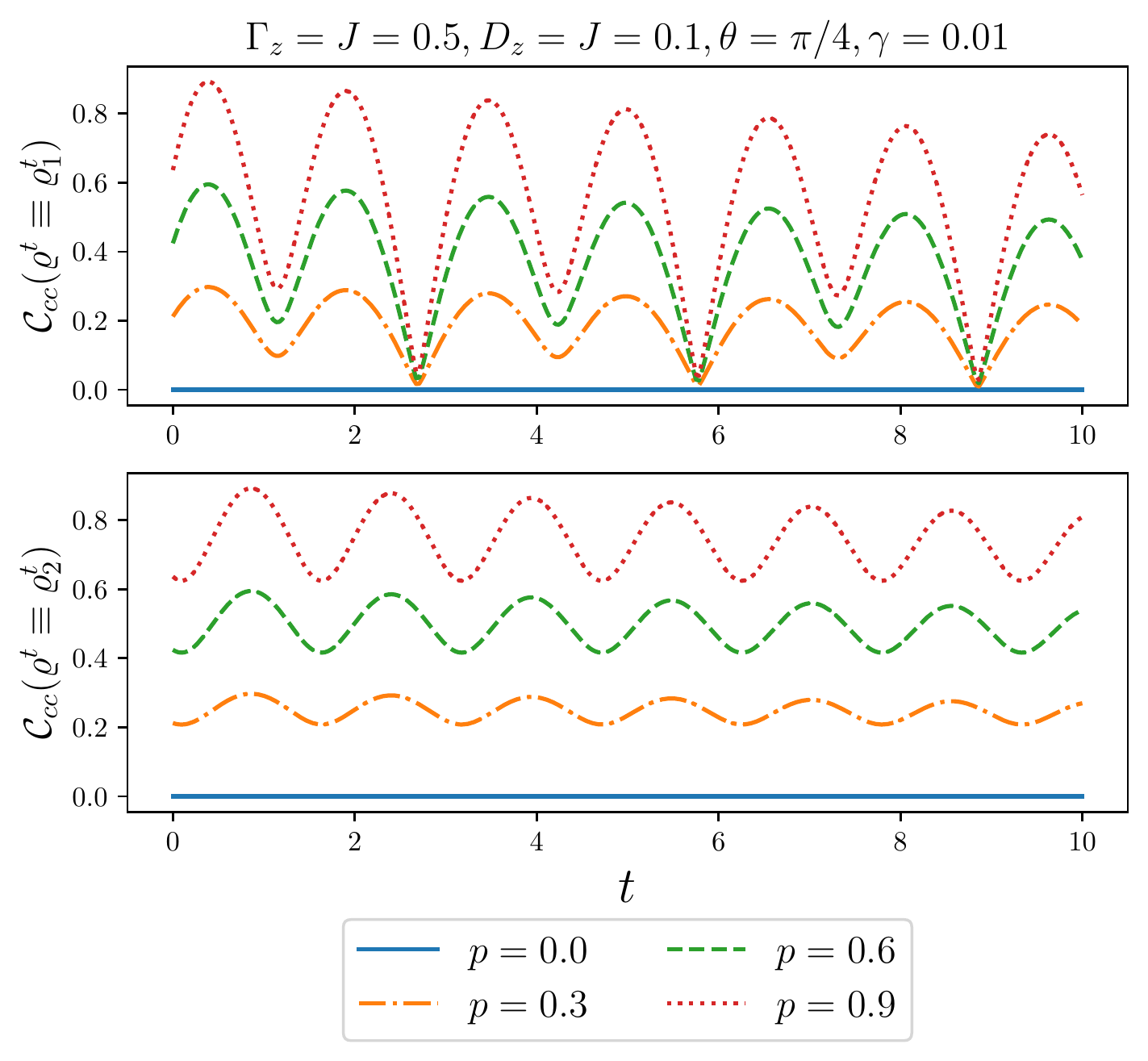}}
\subfigure[]{\label{figure5b}\includegraphics[scale=0.6]{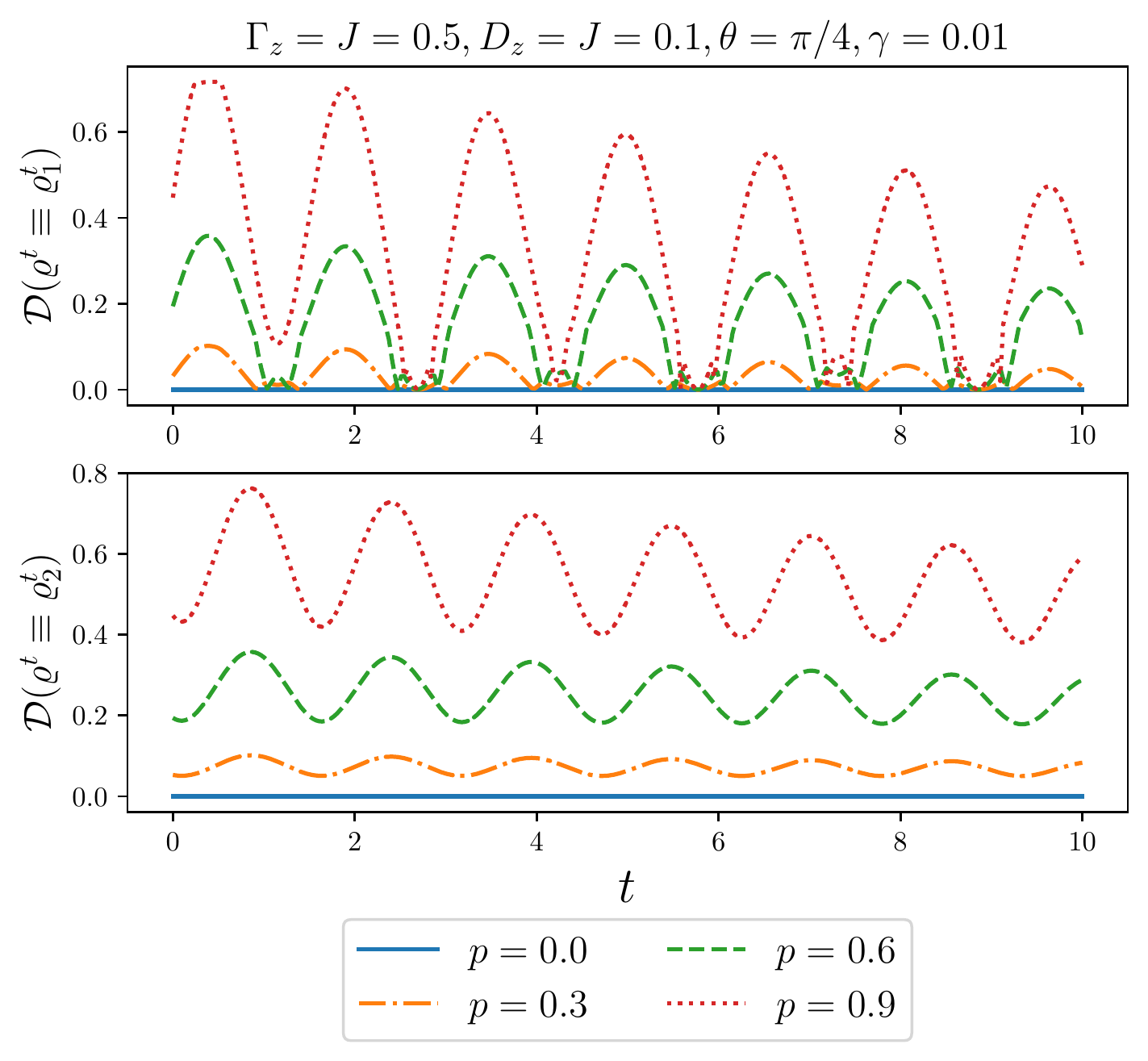}}
\caption{Time-evolution of coherence \ref{figure5a} and quantum correlations \ref{figure5b}. Case 1 ($\varrho^t \equiv \varrho^t _1$) and case 2 ($\varrho^t \equiv \varrho^t _2$), versus the level of purity $p$.}
\label{figure5}
\end{figure}
\end{minipage}
\end{widetext}
In Fig. \ref{figure5}, we show the impact of the level of purity $p$ in the initial state on correlated coherence and quantum discord for both considered states $\varrho_{1}^{t}$ and $\varrho_{2}^{t}$. When $p=0$, the system presents zero coherence and zero discord since the initial state (\ref{wer}) reduces to $\varrho^{t=0}=\varrho_{1}^{t=0}=\varrho_{2}^{t=0}=\frac{1}{4}I_{4}$ which is an incoherent separable state. As it is shown in Figs. \ref{figure5a}-\ref{figure5b}, the evolved density matrix corresponding to $p=0$ does not contain discord-type correlations (${\mathcal D}(\varrho_{1}^{t})={\mathcal D}(\varrho_{2}^{t})=0$) nor correlated coherence ($\mathcal{C}_{cc} (\varrho_{1}^{t})=\mathcal{C}_{cc} (\varrho_{2}^{t})=0$). Raising the level of purity, $p$, improves quantum correlations and coherence within the system, whereas it does not affect the frequency of their oscillations. Both quantifiers, given both cases $\varrho^{t=0}=\varrho_{1}^{t=0}$ and $\varrho^{t=0}=\varrho_{2}^{t=0}$, display damped oscillations. Nevertheless, we recognize in the first case the collapse and revival phenomenon, which is not manifested when $\varrho^{t=0}=\varrho_{2}^{t=0}$. As mentioned earlier, the initial amounts of quantum discord and correlated coherence in the system are those of the initial state given, so it is only natural that they are exclusively dependent on the purity level $p$ and the angle $\theta$. In all figures, it can be seen that the values of quantum coherence and correlations, for both considered cases, are the same before the system starts to evolve (at $t=0$). This is because both Bell-like states have the same superposition and the same amount of entanglement. Furthermore, correlated coherence is always more significant than the nonclassical correlations measured by QD for both initial states.

\section{Conclusion}\label{sec5}
By utilizing correlated coherence and quantum discord, the dynamics of correlated coherence and quantum discord versus intrinsic decoherence rate are investigated in the two-spin XXZ model under the interplay of two different types of interactions. We found that correlated coherence and quantum discord dynamics depend on all system parameters except the anisotropy coupling $J_z$, which has no effects. The two measures go up and down in a pattern that gets smaller and quieter as time passes. In particular, we find that the initial state affects the system's dynamics significantly because it allows it to avoid specific interactions. We also found that the maximum value of the steady state gets smaller by increasing $B, \Gamma_z$, and $D_z$, but increasing $J$ makes the steady state bigger. We show that it is possible to have more robust quantum resources by engineering an appropriate initial state for the system. Moreover, we found that the exchange of information between the system and the magnetic field causes quantum discord to have irregular oscillations for the case $\rho^t \equiv \rho_1 ^t$. We note that the only part contributing to the correlated coherence dynamic is total coherence because local coherence is always zero, meaning that the reduced density matrices of the subsystems are incoherent states. The results of this work give a more in-depth understanding of the influence of specific interactions and initial state and their consequences on the coherence and discord dynamics of the considered quantum system.
\section*{Acknowledgement}
Z. D expresses special thanks to the Abdus Salam International Centre for Theoretical Physics (ICTP) for the hospitality and for providing access to their research facilities during his visit, which helped accomplish some of this work. M. O acknowledges the financial support received from the National Center for Scientific and Technical Research (CNRST) under the Program of Excellence Grants for Research.

\bibliography{references}
\bibliographystyle{ieeetr}

\end{document}